\documentstyle[prb,twocolumn,aps]{revtex}
\tighten
\begin{document}

\title{Dynamic instabilities in resonant tunneling induced by a
magnetic field}
\author{P. Orellana$^{1}$, E. Anda$^{2}$ and F. Claro$^{3}$ }
\address{$^{1}${\it Departamento de F\'{\i}sica,
                     Universidad Cat\'olica del Norte , \\
            Angamos 0610, Casilla 1280, Antofagasta , Chile   }\\
$^{2}${\it Departamento de F\'{\i}sica, Pontificia Universidade Catolica do Rio de Janeiro,\\
Caixa Postal 38071, 22452-970-Rio de Janeiro-Brasil} \\
$^{3}${\it Facultad de F\'{\i}sica, Pontificia Universidad
 Cat\'{o}lica de Chile, \\ Vicu\~{n}a Mackenna 4860, Casilla 306,
 Santiago 22, Chile}}
\address{}
\address{\mbox{ }}
\address{\parbox{14cm}{\rm \mbox{ }\mbox{ }\mbox{ }
We show that the addition of a magnetic field parallel to the
current induces self sustained intrinsic current oscillations
in an asymmetric double barrier structure. The oscillations are
attributed to the nonlinear dynamic coupling of the current to the charge
trapped in the well, and the effect of the external field over
the local density of states across the system. Our results show that
the system bifurcates as the field is increased, and may transit to chaos
at large enough fields.
}}
\address{\mbox{ }}
\address{\parbox{14cm}{\rm PACS numbers:  73.40.Gk,73.40.-c,72.15.Gd}}

\maketitle

\makeatletter

\global\@specialpagefalse
\def\@oddhead{\underline{Accepted in Physical Review Letters\hspace{281pt}
arch-ive/970613}}
\let\@evenhead\@oddhead
\makeatother

\narrowtext

Since the first observation of resonant tunneling through
a semiconductor double barrier
structure (DBS) by Chang, Esaki and Tsu, devices based on this structure have been
found to exhibit a variety of interesting
physical properties\cite{tsu}. Besides a peak due to simple resonant
transmission, structures arising from phonon \cite{gold1} and plasmon
-assisted \cite{zhang}
resonant tunneling, as well as from Landau level matching
\cite{zas,zas1}, have been observed.
Moreover, an intrinsic dynamical bistability and
hysteresis in the negative differential resistance region (NDR) of
the current-voltage characteristic has
been reported \cite{zas,gold,sollner}.
This is a non-linear effect produced by the Coulomb repulsion
experienced by the incoming electrons from the charge buildup in the
space between the barriers\cite{gold}.

Qualitative arguments given by Ricco and Azbel\cite{azbel}
suggest the existence of intrinsic oscillations in a DBS
for the one dimensional case.
When the energy of the incoming electrons
matches the resonance energy
they enter the device and charge the potential well, lifting its bottom,
thus driving the system away from resonance. The ensuing current
decrease lowers the charge in the well bringing the system back
to resonance, and a new cycle in an oscillatory behavior begins.
Numerical calculations support
this prediction for ballistic electrons provided they are not monoenergetic,
and in fact their energy spread
is comparable, or wider than the resonance width \cite{capasso}.

In this letter we report the presence of sustained intrinsic
oscillations when the source is a Fermi sea of
electrons, as one normally has in DBS devices. The oscillations
may be regular or irregular, and are made possible by the presence
of a magnetic field
applied in the direction of the current. They arise thanks to the drastic
change that the field causes on the density of states,
and from the magnetic field induced
enhancement of the nonlinearity of the system, as described below.

Consider a DBS device. In order to study its time evolution under a bias we adopt a
first-neighbors tight-binding model for the Hamiltonian.
Inclusion of a magnetic field B in the growth direction, henceforth called the
z direction, is simple if a parabolic energy dispersion parallel
to the interfaces is assumed.
The field quantizes the motion of the electrons in the $xy$ plane, giving rise
to Landau levels with energies $\epsilon_{n}=(n
+1/2)\hbar\omega_{c}$, where $n=0,1,2,..$ is the Landau index
and $\omega_{c}=eB/m^{*}c$ is the cyclotron frequency.
To a good approximation the longitudinal degrees of freedom
are decoupled from the transverse motion and may be treated
independently. The probability amplitude $b_{j}^{n\alpha}$ for an electron in
a time dependent state $|\alpha>$ to
be at plane j along z and in the Landau level $n$
is determined by the equation of motion

\begin{eqnarray}
i\hbar \frac{db_{j}^{n \alpha}}{dt}&=&(\epsilon_{j}+\epsilon_{n}+
U\sum_{m \beta}|b_{j}^{m \beta}|^{2})b_{j}^{n \alpha}\\ \nonumber
&+&v(b_{j-1}^{n \alpha}+b_{j+1}^{n \alpha}-2b_{j}^{n \alpha}),
\label{eq:time-eq}
\end{eqnarray}

\noindent
where $v$ is the
hopping matrix element between electrons in nearest neighbor
planes, and $\epsilon_{j}$
represents the band contour and external bias.
Here the sum over ($m,\beta$) covers all occupied electron states of
the system,
within the various Landau levels m with  energies
below the Fermi level, incident on the DBS from the emitter side.
In writing Eq. (1) we have
adopted a Hartree
model for the electron-electron interaction, keeping just the intra-atomic
terms as measured by the effective
coupling constant U. We have neglected inter-Landau level
transitions since they are of little importance at not too low
magnetic fields,\cite{yoshio} and averaged over allowed transitions
between degenerate states
within a Landau level, taking advantage of the localization induced
by the gaussian factor in
the Landau basis.
Detailed inclusion of long range effects
make numerical calculations more difficult, without introducing
a qualitative difference in the physical description of the system.
As we will show in what follows, the nonlinear term proportional to U
appearing in the above equation is of key importance in the behavior
of the system.

The time dependent Eq.(1) is solved using a half-implicit
numerical
method which is second-order accurate and unitary  \cite{mains}.
Discretizing the time variable, it can be
written in the form,

\begin{eqnarray}
&\frac{1}{2}&(b_{j-1}^{n\alpha}(t+\delta t) + b_{j+1}^{n\alpha}(t+\delta t)) \\ \nonumber
&+& (\frac{i}{\delta t v} - 1 -2v(\epsilon_{j}+\epsilon_{n}+
V^{Q}_{j}(t)))b_{j}^{n\alpha}(t+\delta t)  = \\ \nonumber
&-\frac{1}{2}&(b_{j+1}^{n\alpha}(t)+b_{j+1}^{n\alpha}(t)) \\ \nonumber
&+& (\frac{i}{\delta t v} + 1 + 2v(\epsilon_{j} + \epsilon_{n} +
V^{Q}_{j}(t)\})b_{j}^{n\alpha}(t),
\label{eq:time-eq2}
\end{eqnarray}

\noindent
where $V^{Q}(t) = U\sum_{m\beta}|b_{j}^{m \beta}(t)|^{2}$ and $\delta t$
is the time increment.
In solving the tridiagonal matrix of Eq.(2), boundary conditions
must be specified at the left ($z=-L$) and right ($z=L$) edges of the structure.
The approach taken here assumes that in the presence of a bias,
the wave function at time $t$ outside the structure is given
 by \cite{mains}

\begin{equation}
b^{n\alpha}_j = Ie^{ik_{\alpha}z_j} + R_{jn}e^{-ik_{\alpha}z_j},\;\; z_j\le -L
\end{equation}

\begin{equation}
b^{n\alpha}_{j}=T_{jn}e^{ik^{\prime}_{\alpha } z_j},\;\; z_j \ge L.
\end{equation}

\noindent
Here $k_\alpha$ and $k^\prime_\alpha=
\sqrt{2m^{*}[\epsilon^{\alpha}-\epsilon_{L}]}/\hbar$ are the wavenumbers
of the incoming and outgoing states, respectively, with
$\epsilon^{\alpha}=\epsilon_n-4 v\; sin^2(k_\alpha a/2)$  the energy
of the incoming particle.
To model the interaction with the particle reservoir outside the structure
the incident amplitude $I$  is
assumed to be a constant independent of the coordinates.
The envelope function of the reflected and transmitted waves, $R_{jn}$
and $T_{jn}$,
are allowed to vary
with $j$, however. Since far from the barriers these quantities are a weak
function of the coordinate $z_j$ we restrict ourselves to the linear
corrections only. This approximation is appropriate provided the time step
$\delta t$ does
not exceed
a certain limiting value. For the results presented here, a maximum value of
$\delta t=3\times 10^{-17} s$ was found sufficient to  eliminate spurious
 reflections at the boundary while maintaining numerical stability
 up to $20\times 10^{-12}s$.

In our numerical procedure the coefficients obtained
for one bias are used as starting point for the
next bias step.
Once $b_{j}^{n\alpha}$ are known, the current at site j is numerically obtained
from \cite{mains}

\begin{equation}
J_j=\frac{e}{\hbar(\pi l_{m})^{2}}\sum_{n}\int_{0}
^{k_{nf}}Im\{b_{j}^{*n\alpha}(b_{j+1}^{n\alpha}-b_{j}^{n\alpha})\}dk_{\alpha},
\end{equation}

\noindent
where $l_{m}=(\hbar c/eB)^{1/2}$ is the magnetic length,
$k_{nf}=\sqrt{2m^{*}(\epsilon_f-\epsilon_n)}/
\hbar$ with $\epsilon_{f}$ the Fermi energy, and the sum is over magnetic
energies $\epsilon_{n}\le \epsilon_{f}$.

We next apply our model to
an asymmetric
$GaAs/AlGaAs$ double barrier structure,
with emitter and collector
barrier thicknesses of $1.12$ $nm$ (2 sites) and $3.36$ $nm$, (6 sites)
respectively, and a well
thickness of $11.2$ $nm$ (20 sites). The second barrier is made wider than
the first in order to enhance the trapping of charge in the well.
For this geometry the first resonance at zero bias and magnetic field
occurs at $30$ $meV$. The conduction band offset is set at $300$ $meV$.
The buffer layers are uniformly doped up to $3$ $nm$ from
either barrier, so as to give a neutralizing free carrier concentration of
$2\times10^{17}cm^{-3}$ at the contacts. In equilibrium and at $B=0$ $T$,
the Fermi level lies $19.2$ $meV$ above the asymptotic
conduction band edge, so that the zero bias resonance lies
well above the Fermi sea.
The parameter values in Eq. (1) are set at
$v= -2.16 eV$ \cite{bastard:wmash} and $U=100 meV$.  The latter was chosen
phenomenologically so as to fit the experimental $I$-$V$ characteristic
for a $GaAs$ devices in the absence
of an external magnetic field.\cite{zas}.
The sample has 400 sites and the
normalization of the wave functions is chosen so that charge from the electrons
filling up to the Fermi energy exactly cancels the positive charge
at the contacts \cite{orella}.
We solved Eq. (2) using the procedure described above,
for an energy mesh
appropriate to compute the integral in Eq.(5). Good convergence was found for
a mesh of 100 points.

For small values of the magnetic field a stationary solution is reached
after a short transient. At large enough bias two stationary
solutions emerge, however, reached separately by either increasing or decreasing
the applied voltage.  In the former case the solution sustains a charged well
while in the latter the well is uncharged and the current is very small.
The effects mentioned
have been observed experimentally\cite{zas1} and are presently
well understood.\cite{orella}

A completely novel feature starts to develop as the magnetic field
is increased. At small values of the external bias a stationary
solution is rapidly reached. As the bias is increased, however, a critical
value arises beyond which no stationary solution exists, and the system
begins to oscillate in a self-sustained way. After a range of voltages
over which the oscillations persist 
a stationary solution is reached again. 
This is illustrated in Fig. 1 where the $I$-$V$ characteristic
for an
applied field of $13 T$ is exhibited. The lobe in the figure shows the current
maxima and minima in the unstable region, its size and location
 depending sensitively on the magnetic field. For the case of Fig. 1 its
 onset is at a rather low bias and reaches into
 the bistable region, while at lower magnetic fields the
 lobe is entirely contained in the low bias range, and lies outside the region
 of bistability.
 In fact, our results show that the well know bistability and the magnetic field
induced instability are entirely
separate phenomena.

In Fig. 2 we show the current at the center
of the well for three different values of the magnetic field and a fixed
bias $V=0.27$ $V$.
For $B=9$ $T$, the
system is seen to reach a stationary state after a transient lasting
about $5ps$ (Fig. 2(a)). For $B=13\; T$ however, the
transient is followed by an oscillation that is never damped out (Fig. 2(b)).
Although not perfectly periodic, the oscillation has two strong
Fourier components at frequencies $\nu\sim 0.3$ and $0.8\; THz$.
The reason why the presence of two frequencies is predominant
remains unclear.
As shown in Fig. 2(c) for $B=17\; T$, at still higher magnetic
fields the oscillation becomes irregular.
A power spectrum
of this latter signal shows that the narrow peaks observed below $1\; THz$
are replaced by a broad low frequency structure suggesting that
a chaotic regime has been reached. Note that the initial condition
from which all three curves were
obtained is the stable solution at $0.2\; V$. When reaching a bias
of $0.27\; V$ from a closer voltage the transient oscillations
are less pronounced, yet their relaxation time is of the same order
as in the more extreme voltage jump exhibited in the figure.
Although the amplitude of the sustained oscillations depends
somewhat on the initial conditions, their frequency and structure does not.

Figure 3 shows
the regions where the various structures we
have described appear in the two parameter space $V$-$B$.
The thick lines delimit the range in which
two stable solutions may exist for the same
applied potential. The dotted regions mark areas where self-sustained
oscillatory solutions appear. Regions with highly irregular oscillations
suggesting a chaotic behavior appear filled with a brick design.
We note that while we include in our simulations
up to several hundred thousands time steps, the number of oscillations
covered before numerical instabilities arise are not enough to obtain a fully
detailed power spectrum at all values of parameters. The
search of stationary solutions using a standard self-consistent iterative loop,
however, did
 show unambiguously a transition to chaos through successive bifurcations
within the region where no stationary solutions exist. Although the
iteration number in this case is not a true time variable, the parameter values
at which non stationary solutions exist and the nature of the latter
coincide with those obtained in the true-time dependent
analysis. We used this fact in sketching Fig. 3.

We interpret our findings in the following way. In general, current
flows through the system as long as a tunneling resonance
lies within the emitter Fermi sea. The resonance
acts effectively as an energy filter for the transmitted electrons.
We assume that at very low bias it lies above the Fermi energy
so that no current flows. As the bias increases the resonance drops,
reaching eventually
the Fermi energy, thus opening a channel for electrons to tunnel
through the double barrier. A
current is thus established. What happens next depends sensitively
on the presence or absence of a magnetic field.
In the absence of the electron-electron interaction the current,
roughly speaking,
grows linearly with the bias if B=0. This behavior
is due to the gradual
increase in tunneling
states that satisfy the conservation laws as the voltage is raised.
At a critical bias the falling resonance
reaches the bottom of the conduction band, after which the current
drops abruptly. This overall behavior gives the I-V curve its
characteristic triangular shape.
For finite B however, also in the absence of electron-electron interaction,
owing to the new density of states each time
a Landau level enters the Fermi sea the current rises abruptly,
remaining essentially constant until the next level comes in.
The I-V curve acquires then a staircase shape, with a step rise
made smooth by the resonance line shape.
When the resonance reaches the bottom of the first Landau
level in the emitter, the current falls abruptly, as in the B=0 case.

We next take into account the electron-electron interaction.
As the resonance enters the Fermi sea, current begins to
flow and
charge is trapped in the potential well, rising
its bottom and pushing the resonance towards higher energies.
The current drops,
some of the accumulated charge leaks out allowing the current to flow more
easily once again, and a new cycle begins. When B=0 this process
occurs at very low current, and as we have
verified numerically, the ensuing oscillations are
damped out in all cases.
At finite field however, as the resonance enters the Fermi sea,
the abrupt rise of the current to a large value
triggers oscillations that may be sustained by the feedback
mechanism if the field is large enough.

The transition from
damped to sustained oscillations may be appreciated in Fig. 2.
As the magnetic field is raised, the degeneracy of the
resonance increases thus allowing more charge to be
trapped between the barriers. The nonlinear coupling
in Eq. (1) thus grows with the field, making the latter an effective
tunable parameter for the amount of nonlinearity in the system.
Note also that while the dynamic bistability characteristic
of DBS is related to the drop of the resonance below the bottom of the
Fermi sea, the effect we are discussing arises as the resonance
enters the Fermi sea, in support of our numerical finding
that they are entirely independent effects.

According to the picture drawn above
the charge in the well lags the current, as exhibited in the inset
of Fig. 2(b). Here the charge at the
center of the well (dashed line) is plotted together with the current at the same
point (full line), the former displaced a time $\tau \sim 1.4 ps$
to the left with respect to the
latter. The relaxation time $\tau$ and the period
of the oscillations are determined by the tunneling time
for the electrons to leak out through the barriers and may be adjusted
by modifying the barrier thicknesses.
The oscillatory current is always non-uniform accross the device,
with prominent oscillations in the
well and the emitter side, while the collector current exhibits
a weaker structure. 
This is due to the large width of the collector barrier relative to that
in the emitter side,
a necessary feature of design
to have enough charge accumulation in the well, and thereby, a strong
electron-electron interaction.
In fact, nonlinear effects disappear in our sample if the width of the collector
barrier is reduced from six to just three sites, in agreement with 
experimental observations. \cite{eaves}
The very high frequency
oscillations seen early in the process and shown in the inset
of Fig. 2(c), are quickly damped out and do not contribute to the
long term behavior of the system. We attribute them to a transient pulse
going back and forth in the space between the barriers before it looses
coherence and decays. Its period corresponds to electrons at the Fermi
velocity being successively reflected by the barriers in
the well region.

There is a wide region of accessible values of bias and
magnetic field where oscillatory currents are either present or absent.
This opens up the interesting and unique
possibility of studying experimentally the transition
from stationary to oscillatory behavior and possibly chaos, as the
magnetic field is increased in an asymmetric $DBS$ device under bias.
When oscillating, the system could act as an electromagnetic
generator in the $THz$ region. Research incorporating the interaction between
the current and the radiation field is currently under progress.

Work supported by FONDECYT
grants 1950190, 3950026 and 1960417, Fundaci\'on Antorchas/Vitae/Andes grant
B-11487/4B005,
 CNPq and FINEP.

\newpage

\begin{figure}
        \caption {$I$-$V$ characteristic of the asymmetric $DBS$ at $B=13$ $T$.
        The lobe edges mark the oscillation extremes when
        an unstable solution is present. The dashed line represents the lower
        arm of the bistable regime, reached when the bias is being decreased.}
\label{fig:fig1}
\end{figure}

\begin{figure}
        \caption{Current as a function of time at the center of
the well for $V=0.27$ $V$ and an applied magnetic field of (a) 9T,
(b) 13T and (c) 17T. The inset in (b) shows the charge and
current shifted one relative to the other by 1.4 ps. The inset
in (c) exhibits a detail of the high frequency transient
oscillations.}
\label{fig:fig2}
\end{figure}

\begin{figure}

  \caption { Phase diagram in the $B$,$V$ parameter space. The
  thick lines delimit the region where bistability occurs.
  Dotted areas correspond to oscillatory behavior,
  while the brick design marks
 the regions where highly irregular solutions are present. The filled
 circles mark the values used in Fig. 2}
\label{fig:fig3}
\end{figure}


\begin{thebibliography}{20}
\bibitem{tsu} L. L. Chang, L. Esaki and R. Tsu, Appl. Phys. Lett.
{\bf 24}, 593 (1974)
\bibitem{gold1} V. J. Goldman, D. C. Tsui and J. E. Cunningham,
Phys. Rev. {\bf B36}, 7635 (1987)
\bibitem{zhang} C. Zhang, M. L. F. Lerch, A. D. Martin, P. E. Simmonds and
L. Eaves, Phys. Rev Lett.{\bf 72}, 3397 (1994)
\bibitem{zas} A. Zaslavsky, V. J. Goldman, D. C. Tsui and J. E. Cunningham,
Appl. Phys. Lett. {\bf 53}, 1408 (1988)
\bibitem{zas1} A. Zaslavsky, D. C. Tsui, M. Santos and M. Shayegan,
Phys. Rev. {\bf B40}, 9829 (1989)
\bibitem{gold} V. J. Goldman, D. C. Tsui and J. E. Cunningham,
Phys. Rev. Lett. {\bf 58}, 1256 (1987)
\bibitem{sollner} T. C. L. G. Sollner, Phys. Rev. Lett. {\bf 59}, 1622 (1987)
\bibitem{azbel} B. Ricco and M. Ya. Azbel,
Phys. Rev. {\bf B29}, 1970 (1984)
\bibitem{capasso} G. Jona-Lasinio, C. Presilla and F. Capasso,
Phys. Rev. Lett. {\bf 68}, 2269 (1992)
\bibitem{yoshio} D. Yoshioka and P. A. Lee, Phys. Rev. {\bf B27}, 4986 (1983)
\bibitem{mains} R. Mains and G. Haddad, J. Appl. Phys {\bf 64}, 3564
(1988)
\bibitem{bastard:wmash} G. Bastard, in: {\it Wave Mechanics Applied to Semiconductor Heterostructures
 (Les \'editions de physique, 1988)} p. 34.
\bibitem{orella} P. Orellana, F. Claro, E. Anda and S. Makler,
Phys. Rev. {\bf B 53} (1996) 12967
\bibitem{eaves} L. Eaves, M.L Leadbeater and C.R.H. White, Physica {\bf
B} 263 (1991)

\end{thebibliography}
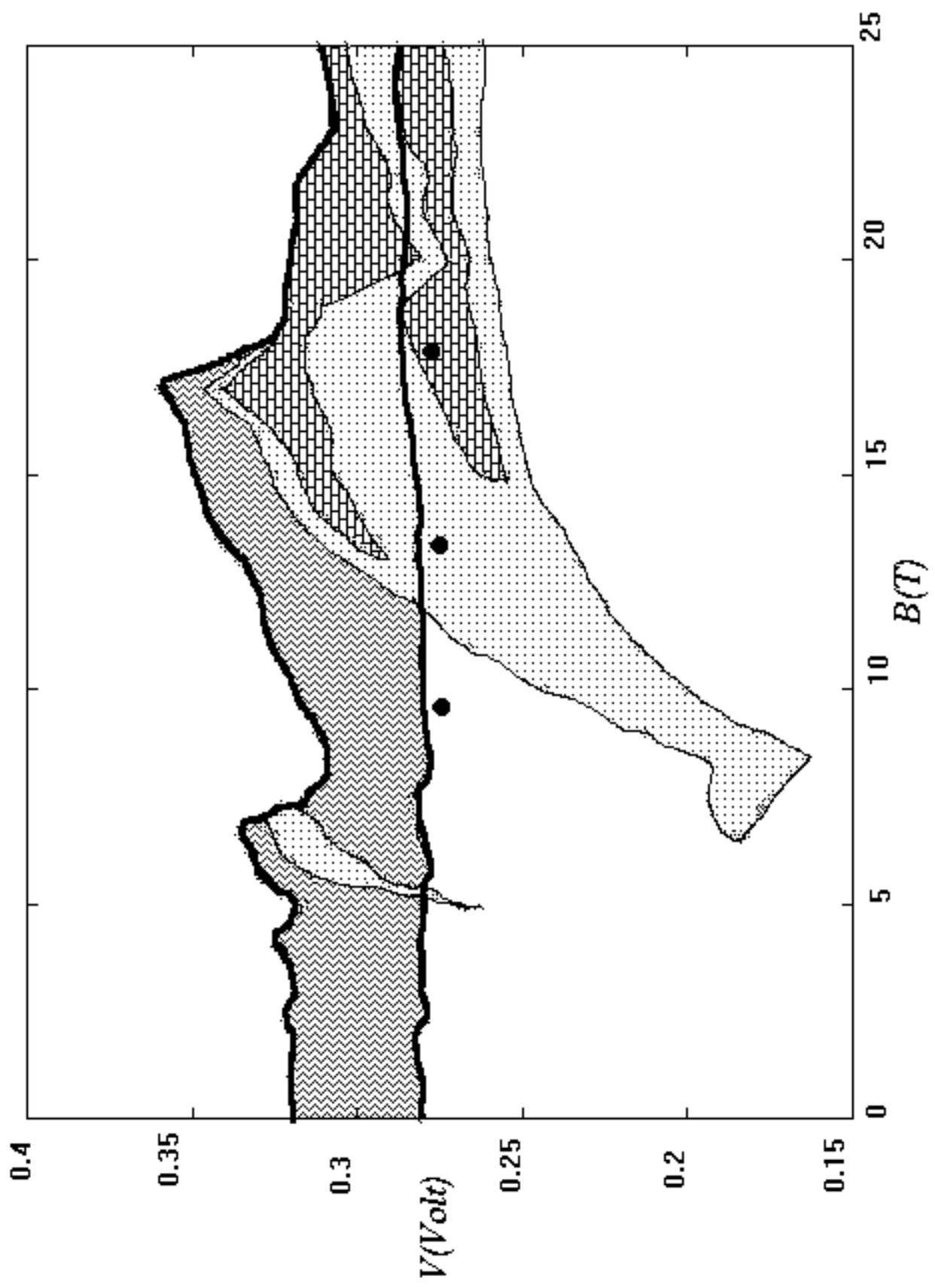
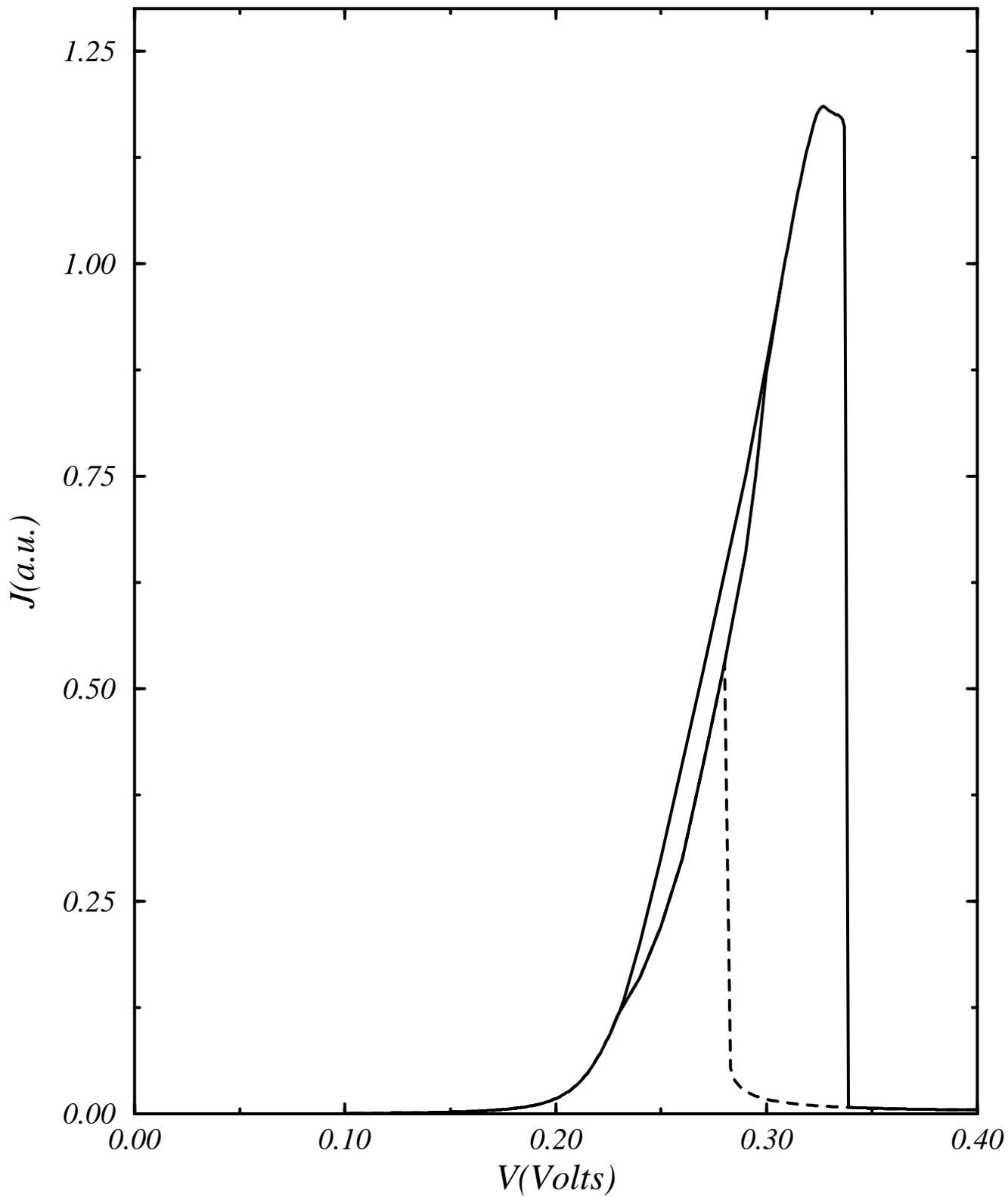
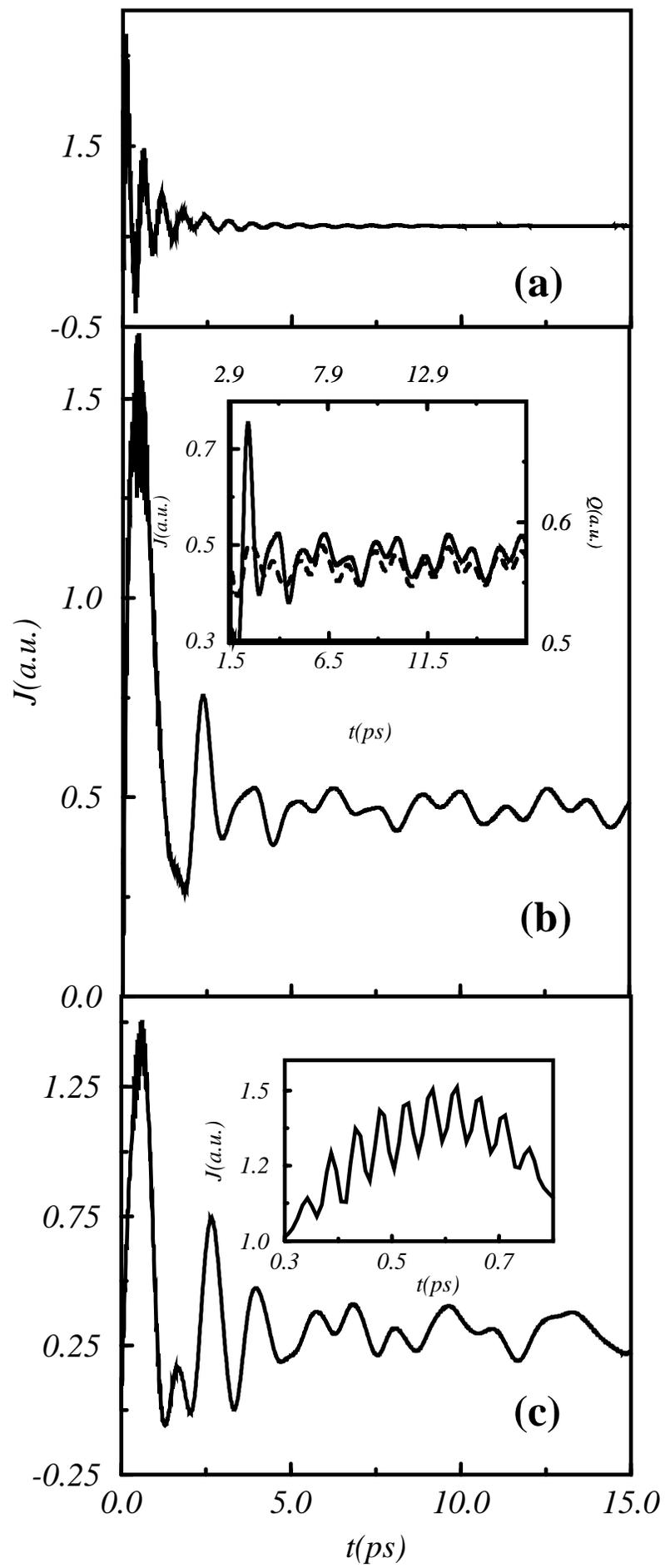
\end{document}